\documentclass[preprint,12pt]{elsarticle}

\usepackage{amssymb}
\usepackage{amsmath}
\numberwithin{equation}{section}

\journal{Physica A}

\begin{document}

\begin{frontmatter}



\title{On the statistical distributions of substance moving through the nodes of a channel of  network}


\author[1,2]{Nikolay K. Vitanov\corref{cor1}}
\author[1]{Kaloyan N. Vitanov}
\cortext[cor1]{corresponding author, e-mail:vitanov@imbm.bas.bg}

\address[1]{Institute of Mechanics, Bulgarian Academy of Sciences, Acad.
G. Bonchev Str., Block 4, 1113 Sofia, Bulgaria }
\address[2]{Max-Planck Institute for the Physics of Complex Systems, Noethnitzerstr.
38, 0187 Dresden, Germany}

\begin{abstract}
We discuss a model of motion of substance through the nodes of a  channel of a network. 
The channel can be modeled by a chain of urns where each urn can exchange substance with the neighboring  
urns. In addition the urns can exchange substance with the network nodes and the new point is that we include in the model the possibility for exchange of substance among the urns (nodes)  and the environment of the network. 
We consider  stationary regime of motion of substance through a finite channel (stationary regime of 
exchange of substance along the chain of urns) and obtain
a class of statistical distributions of substance in the nodes of the channel. 
Our attention is focused on this class of distributions and we show that for the case of finite channel
the obtained class of distributions   contains as particular cases truncated versions of the families of distributions
of Katz, Ord, Kemp, etc.
The theory for the case of infinite  chain of urns is presented in the Appendix.
\end{abstract}


\end{frontmatter}



\section{Introduction}
Research on the mathematical models of  complex systems intensified much in the
last years. Just several examples are connected to networks \cite{x32}, \cite{x33};
population dynamics \cite{x2} - \cite{x4}; biology and physiology \cite{x12} - \cite{x26}; social dynamics \cite{x28}, \cite{x30} ; etc. \cite{nfirst} - \cite{nlast}.
Models of flows in networks are widely used in the study of various kinds of 
problems, e.g, flows in computer networks \cite{bing}, flows in financial networks \cite{eboli}, flows in
electrical and communication networks \cite{ch2}, transportation problems \cite{ff}-\cite{ch1}, etc.. 
At the beginning  the research  was  focused on problems such as 
maximal flows in a network.
Then the field of research  expanded to:  shortest path finding, 
self-organizing network flows,  facility layout and location,  modeling and optimization of scalar flows in networks \cite{ambro},  optimal  electronic route guidance in urban traffic networks \cite{hani}, isoform identification of RNA \cite{bernard},  memory effects \cite{rosvall}, etc. (see, e.g., \cite{masson} - \cite{skutella}). Below we shall discuss the motion of substance in a channel of a
network on the basis of a discrete - time model in presence of possibility for exchange of
substances: (i) between the channel and the network and (ii) between the channel and the environment of the network. The discussed model  extends the model studied in \cite{vk1} and has many potential applications, e.g., (i) to 
model flow of a substance through a channel and use of part  of the substance in some industrial process in the nodes of the channel or (ii)  to model  human  migration flows.  The application (ii) is important as  the probability and deterministic
models of human migration are interesting  from the point of view of applied mathematics
\cite{will99} - \cite{grd05} and the  human migration flows are very important for
taking decisions about economic development of regions of a country \cite{everet} -
\cite{borj}.  Human migration is studied in connection with , e.g.,: (i) 
migration networks \cite{vk1}, \cite{fawcet} - \cite{vk1x}; (ii) ideological struggles 
\cite{vit1}, \cite{vit2} ; (iii)  waves and statistical distributions in 
population systems \cite{vit3} - \cite{vit6}. Models, similar to human migration models
are used also in other research areas \cite{sg1}, \cite{vb}. 
We shall connect below a class of flows in a channel of network to a class 
of statistical distributions occurring in a chain of urns, i.e., we shall discuss also an appropriate
urn model of the studied network flow. This is an interesting connection as many important results in probability theory may be derived from urn models and  the urn models are much
applied in the research on various problems, e.g., from genetics, clinical trials, biology, social dynamics, military theory, etc. \cite{jkotz} - \cite{gardy}.
\par 
The text of the paper  is organized as follows. 
\begin{itemize}
	\item Sect.2:  we formulate the model of motion of substance 
in a finite channel of nodes in a network. The new point is that in the model equations terms are included that
account for the possibility for exchange of substance among the nodes of the channel
and the environment of the network. 
\item Sect.3:
we discuss distributions connected to the stationary regime of motion of substance through the channel
of nodes (stationary regime of motion of substance through the chain of urns). 
\item Sect. 4: Several 
concluding remarks. 
\item
Appendix A contains the description of the case of infinitely
long channel of nodes (infinite number of urns in the chain of urns) that is interesting also from the 
point of view of the probability theory.
\end{itemize}
\section{Formulation of the model}
We study a network consisting of nodes connected by edges. In more detail
we consider  a channel in this network and the channel can be modeled by chain of urns  
Each urn of the chain of urns is placed at a node of the network.
The urns are connected, i.e.,  one can take substance  from one
of the urns and can put this substance in the previous or in the next urn from
the chain of urns. There can be also exchanges: (i)  between the urns and the network
and (ii) between an urn and the environment of the network.
\begin{figure}[!htb]
\centering
\includegraphics[scale=.5]{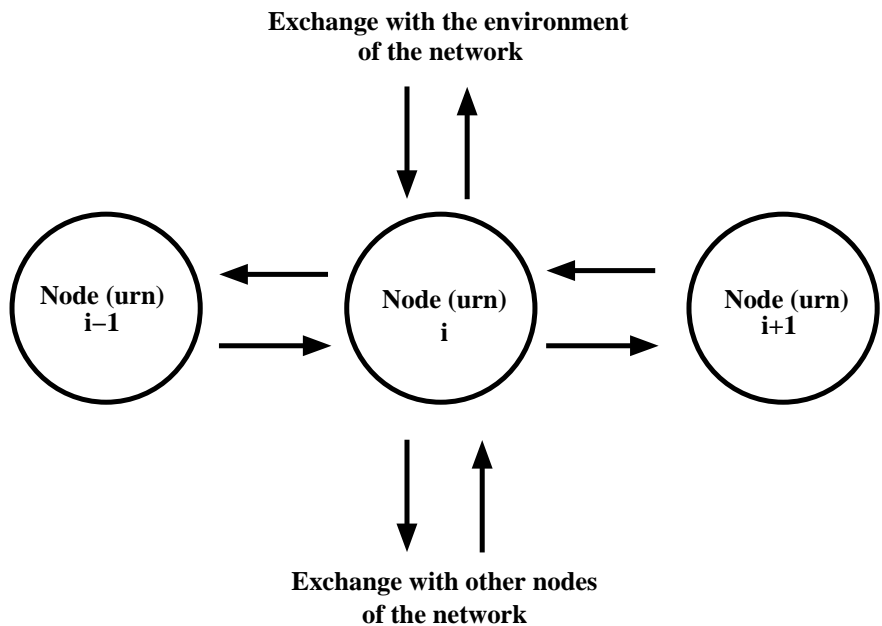}
\caption{ Processes connected with the $i$-th urn of the chain of urns
$i=0,\dots,N$. Note that the urns with numbers $0$ and $N$ are connected only to one
of the other urns. These tho urns may exchange substance only with
one of the other urns.
}
\end{figure}
The possible exchange processes for the $i$-th urn are shown in Fig. 2.  The $i$-th urn
can  exchange  substance with the $(i-1)$-th and with the $(i+1)$-st urn. The
$i$-th urn can also exchange  substance with the network nodes outside the channel
 and with the environment of the network. Let us denote as: (i) "leakage":
the process of motion of substance from the urn to a node of the network or to the environment of the network; (ii)
"pumping":   the process of motion of substance from the network or from the environment
of the network to the  urn. 
\par
We assume further  that  the sequence of urns consists  of a chain 
of $N+1$ urns (labeled  from $0$ to $N$) connected by corresponding edges (ways for motion of 
the substance). Each edge connects two urns and each urn is connected to two 
edges except for the $0$-th urn and $N$-th urn that are connected to one edge. We shall 
consider  discrete time and let us assume below that the time intervals have
equal length.  At each time interval the substance in an urn of the chain of urns 
can participate in one of the following four processes: 
\begin{description}
\item[(1.)] the substance remains in the same urn; 
\item[(2.)] the substance moves to the previous or to the next urn 
(i.e., the substance may move from  the urn $m$ to the urn $m+1$ or from the urn $m$ to
the urn $m-1$); 
\item[(3.)]  the substance''leaks'' from the 
urn $m$: this means that the ''leaked'' substance does not belong anymore to 
the sequence of urns. Such substance may spread through the network or through environment of the 
network. Thus we have two kinds of ''leakage": (i) leakage from the urn $m$ to nodes of  the network and (ii)
leakage from the urn $m$ to the environment of the network; 
\item[(4.)] the substance is''pumped'' to the 
urn $m$. Two kinds of ''pumping" are possible: (i) pumping from the network to the urn $m$ and (ii)
pumping to from the environment of the network to the urn $m$ ; 
\end{description}
\par
Let us formalize mathematically the above considerations. Before starting this we note that
a particular case of the model formulated here was discussed in \cite{vk1} but for convenience 
of the reader we shall write the entire model  below as the final equations are more complicated that the
model equations from \cite{vk1} . 
\par
We consider discrete time $t_k$, $k=0,1,2,\dots$ and  denote the amount of
the substance  in the $i$-th urn of the chain at the beginning of the
time interval $[t_k, t_k + \Delta t]$ as $x_i(t_k)$. For the processes happening in this
time interval in the $n$-th urn of the chain we shall use the following 
notations:
\begin{description}
\item[1.]
$i^e_n(t_k)$ and $o^e_n(t_k)$ are the amounts of inflow and outflow   of 
substance  from the environment to the $n$-th urn  of the chain (the 
upper index $e$ denotes that the quantities are for the environment);  
\item[2.] 
$o^c_n(t_k)$ is the amount of outflow of substance from the $n$-th urn of the
chain to the $(n+1)$-th urn of the chain (the upper index $c$ 
denotes that the quantities are for the chain of urns); 
\item[3.]
$i^c_n(t_k)$ is the amount of the inflow of substance from the $(n+1)$ urn of 
the chain to the $n$-th urn of the chain; 
\item[4.]
$o^n_n(t_k)$ and $i^n_n(t_k)$ are the amounts of outflow  
and inflow of substance  between the $n$-th urn of the chain and  
the network (the upper index $n$ 
denotes that the quantities are for the network).
\end{description}
For  the $0$-th urn we have exchange of 
substance with the environment (inflow and outflow); exchange of substance 
with the next urn of the chain (inflow and outflow) and exchange (inflow and outflow) of 
substance with the network. Thus the change  of the amount of substance in the $0$-th 
node of the channel is described by the relationship
\begin{equation}\label{eq1}
x_0(t_{k+1}) = x_0(t_k)+ i^e_0(t_k) - o^e_0(t_k) - o^c_0(t_k) + i^c_0(t_k) 
-o^n_0(t_k) + i^n_0(t_k)
\end{equation}
For the urns of the chain  numbered by $i=1,\dots,N-1$ there is exchange
of substance with the environment, exchange of substance with the network  and exchange
of substance with $(i-1)$-st and $(i+1)$-st urns of the chain of urns. Thus the change  of the amount 
of substance  in the $i$-th urn of the channel  is described by the relationship
\begin{eqnarray}\label{eq2}
x_i(t_{k+1}) = x_i(t_k)+ i^e_i(t_k) - o^e_i(t_k)  +o^c_{i-1}(t_k)
-i^c_{i-1}(t_k) - o^c_i(t_k) + i^c_i(t_k) - \nonumber \\
o^n_i(t_k) + i^n_i(t_k) , \ \ \ 
i=1,\dots,N-1 \nonumber \\
\end{eqnarray}
For the $N$-th urn of the channel there is exchange of substance with the
environment,  exchange of substance with the network and exchange of substance  
with $(N-1)$-st urn of the chain of urns. Thus the change  of the amount of 
substance in the $N$-th urn  of the chin of urns  is described by the relationship
\begin{eqnarray}\label{eq3}
x_N(t_{k+1}) = x_N(t_k)+ i^e_N(t_k) - o^e_N(t_k)  + o^c_{N-1}(t_k)
-i^c_{N-1}(t_k) - \nonumber \\
o^n_N(t_k) + i^n_N(t_k) 
\end{eqnarray}
\par
Eqs.(\ref{eq1}) - (\ref{eq3}) describe the general case of motion of substance
along a chain of urns connected to a network and to the environment of this network. 
The addition with respect to \cite{vk1} is that we shall allow below an exchange of substance
between and urn and environment of the network. Thus we will  obtain statistical
distributions that contain as particular cases   the statistical distributions studied in \cite{vk1}.
\par
We shall continue our study  by consideration of  the following particular case of the quantities from
the system of equations (\ref{eq1}) - (\ref{eq3}):
\begin{itemize}
\item Exchange between the chain of  urns and the  environment of the network
\begin{eqnarray}\label{pc1}
i^e_0(t_k) &=& \sigma_0(t_k) x_0(t_k); \ \ o^e_0(t_k) = \mu_0(t_k) x_0(t_k) \nonumber \\
i^e_i(t_k) &=& \sigma_i(t_k) x_i(t_k); \ \ o^e_i(t_k) = \mu_i(t_k) x_i(t_k) , \ i=1,\dots, N-1 \nonumber \\
i^e_N(t_k) &=& \sigma_N(t_k) x_N(t_k); \ \ o^e_N(t_k) = \mu_N(t_k) x_i(t_k) 
\end{eqnarray}
\item Exchange between the chain of urns and the network
\begin{eqnarray}\label{pc2}
i^n_0(t_k) &=& \epsilon_0(t_k) x_0(t_k); \ \ o^n_0(t_k) = \gamma_0(t_k) x_0(t_k) \nonumber \\
i^n_i(t_k) &=& \epsilon_i(t_k) x_i(t_k); \ \ o^n_i(t_k) = \gamma_i(t_k) x_i(t_k), \ i=1,\dots,N-1 \nonumber \\
i^n_N(t_k) &=& \epsilon_N(t_k) x_N(t_k); \ \ o^n_N(t_k) = \gamma_N(t_k) x_N(t_k)
\end{eqnarray}
\item Exchange within the chain of urns
\begin{eqnarray}\label{pc3}
o^c_0(t_k) &=& f_0(t_k) x_0(t_k); \ i^c_0(t_k) = \delta_1(t_k) x_1(t_k) \nonumber \\
o^c_i(t_k) &=& f_i(t_k) x_i(t_k); \ i^c_i(t_k) = \delta_{i+1}(t_k) x_{i+1}(t_k), \ i = 1,\dots, N-2 \nonumber \\
o^c_{N-1}(t_k) &=& f_{N-1}(t_k) x_{N-1}(t_k); \ i^c_{N-1}(t_k) = \delta_{N}(t_k) x_{N}(t_k)
\end{eqnarray}
\end{itemize}
\par 
For this particular case the system of equation (\ref{eq1}) - (\ref{eq3}) becomes
\begin{eqnarray}\label{ex1}
x_0(t_{k+1}) = x_0(t_k) + 
\sigma_0(t_k) x_0(t_k) - \mu_0(t_k) x_0(t_k) - f_0(t_k) x_0(t_k) + \nonumber \\
\delta_1(t_k) x_1(t_k) - \gamma_0(t_k) x_0(t_k) + \epsilon_0(t_k) x_0(t_k)
\end{eqnarray}
\begin{eqnarray}\label{ex2}
x_i(t_{k+1}) = x_i(t_k) + 
\sigma_i(t_k) x_i(t_k) - \mu_i(t_k) x_i(t_k) + f_{i-1}(t_k) x_{i-1}(t_k) - \nonumber \\
\delta_{i}(t_k) x_{i}(t_k) - f_i(t_k) x_i(t_k) + \delta_{i+1}(t_k) x_{i+1}(t_k) -
\gamma_i(t_k) x_i(t_k) + \epsilon_i(t_k) x_i(t_k)  , \nonumber \\
i=1,\dots,N-1 \nonumber \\
\end{eqnarray}
\begin{eqnarray}\label{ex3}
x_N(t_{k+1}) = x_N(t_k)+ 
\sigma_N(t_k) x_N(t_k) - \mu_N(t_k) x_i(t_k) + f_{N-1}(t_k) x_{N-1}(t_k) - \nonumber \\
\delta_{N}(t_k) x_{N}(t_k) - \gamma_N(t_k) x_N(t_k) + \epsilon_N(t_k) x_N(t_k)
\nonumber \\
\end{eqnarray}
Below we shall  consider the case of absence of inflow from $i+1$-st urn to the $i$-th urn (no
flow of substance  in the direction opposite to the direction from the $0$-th urn to the
$N$-th urn of the chain of urns). In this case the system of equations (\ref{ex1}) - (\ref{ex3})
becomes
\begin{eqnarray}\label{ey1}
x_0(t_{k+1}) = x_0(t_k) + 
\sigma_0(t_k) x_0(t_k) - \mu_0(t_k) x_0(t_k) - f_0(t_k) x_0(t_k) - \nonumber \\
\gamma_0(t_k) x_0(t_k) + \epsilon_0(t_k) x_0(t_k)
\end{eqnarray}
\begin{eqnarray}\label{ey2}
x_i(t_{k+1}) = x_i(t_k) + 
\sigma_i(t_k) x_i(t_k) - \mu_i(t_k) x_i(t_k) + f_{i-1}(t_k) x_{i-1}(t_k) - \nonumber \\
 f_i(t_k) x_i(t_k) - \gamma_i(t_k) x_i(t_k) + \epsilon_i(t_k) x_i(t_k)  , \nonumber \\
i=1,\dots,N-1
\end{eqnarray}
\begin{eqnarray}\label{ey3}
x_N(t_{k+1}) = x_N(t_k)+ 
\sigma_N(t_k) x_N(t_k) - \mu_N(t_k) x_i(t_k) + f_{N-1}(t_k) x_{N-1}(t_k) - \nonumber \\
 \gamma_N(t_k) x_N(t_k) + \epsilon_N(t_k) x_N(t_k) \nonumber \\
\end{eqnarray}
We note that the system of equations (\ref{ey1}) - (\ref{ey3}) contains as particular case the system of equations
(2.8) - (2.10) from \cite{vk1}. 
We shall study the stationary case of the model equations (\ref{ey1}) - (\ref{ey3}) in more detail
below.
\section{Distributions of substance for stationary regime  of motion of substance  through the chain of urns }
We  consider the case of stationary motion of the substance  through the chain of urns.
In this case $x_i(t_{k+1}) = x_i(t_k)$, $i=0,\dots,N$ and the system of equations (\ref{ey1}) - (\ref{ey3})
becomes
\begin{eqnarray}\label{ez1} 
\bigg[ \sigma_0(t_k) - \mu_0(t_k) - f_0(t_k)  -  \gamma_0(t_k) + \epsilon_0(t_k) \bigg] x_0(t_k) = 0
\end{eqnarray}
\begin{eqnarray}\label{ez2}
\bigg[ \mu_i(t_k) + f_i(t_k) + \gamma_i(t_k) - \sigma_i(t_k) - \epsilon_i(t_k) \bigg] x_i(t_k) = f_{i-1}(t_k) x_{i-1}(t_k),
\nonumber \\
i=1,\dots,N-1
\end{eqnarray}
\begin{eqnarray}\label{ez3}
\bigg[ \mu_N(t_k) + \gamma_N(t_k) - \sigma_N(t_k) - \epsilon_N(t_k) \bigg] x_N(t_k) = f_{N-1}(t_k) x_{N-1}(t_k)
\end{eqnarray}
Below we discuss the model described by Eqs.(\ref{ez1}) -
(\ref{ez3}) for the case when the parameters of the model are time independent
(i.e., when $\sigma_i(t_k) = \sigma$; $\mu_i(t_k) = \mu_i$;
$\gamma_i(t_k) = \gamma_i$; $\epsilon_i(t_k) = \epsilon_i$; $f_0(t_k)=f_0$;
$i=0,\dots,N$. These  parameters do not depend on the time but they may depend on
$i$ and also on other parameters connected to the network and to the environment
of the network. In this case the system of model equations becomes
\begin{eqnarray}\label{eu1} 
\bigg[ \sigma_0 - \mu_0 - f_0  -  \gamma_0 + \epsilon_0 \bigg] x_0 = 0
\end{eqnarray}
\begin{eqnarray}\label{eu2}
\bigg[ \mu_i + f_i + \gamma_i - \sigma_i - \epsilon_i \bigg] x_i = f_{i-1} x_{i-1}, \ \
i=1,\dots,N-1
\end{eqnarray}
\begin{eqnarray}\label{eu3}
\bigg[ \mu_N + \gamma_N - \sigma_N - \epsilon_N \bigg] x_N = f_{N-1} x_{N-1}
\end{eqnarray}
\par 
We obtain from the system of equations (\ref{eu1}) - (\ref{eu3})
the following relationships for the corresponding class of statistical distributions:
\begin{eqnarray}\label{d1}
f_0 &=& \sigma_0 - \mu_0  -  \gamma_0 + \epsilon_0; \ \
x_i = x_0 \prod \limits_{k=1}^i \frac{f_{k-1}}{\mu_k + f_k + \gamma_k - \sigma_k - \epsilon_k } 
\nonumber \\
x_N &=& x_0 \frac{f_{N-1}}{\mu_N + \gamma_N - \sigma_N - \epsilon_N }\prod \limits_{k=1}^{N-1} \frac{f_{k-1}}{\mu_k + f_k + \gamma_k - \sigma_k - \epsilon_k } \nonumber \\
x &=& \sum \limits_{i=0}^N x_i = x_0 \Bigg[ 1 + \sum \limits_{i=1}^{N-1}\prod \limits_{k=1}^i \frac{f_{k-1}}{\mu_k + f_k + \gamma_k - \sigma_k - \epsilon_k } + \nonumber \\
&&\frac{f_{N-1}}{\mu_N + \gamma_N - \sigma_N - \epsilon_N }\prod \limits_{k=1}^{N-1} \frac{f_{k-1}}{\mu_k + f_k + \gamma_k - \sigma_k - \epsilon_k }  \Bigg]
\end{eqnarray}
Eqs.(\ref{d1}) leads to a class of statistical distributions as follows. We have $x_i$ and $x$ and  we can consider the statistical distribution $y_i=x_i/x$ of the 
amount of substance along the urns of the chain of urns. $y_i$ can be considered as 
probability values of distribution of a discrete random variable $\zeta$: $y_i = p(\zeta =i)$, $i=0,\dots, N$. For this distribution we obtain
\begin{eqnarray}\label{d2}
y_0 =  \frac{1}{1 + \sum \limits_{l=1}^{N-1}\prod \limits_{k=1}^l \frac{f_{k-1}}{\mu_k + f_k + \gamma_k - \sigma_k - \epsilon_k } + 
	\frac{f_{N-1}}{\mu_N + \gamma_N - \sigma_N - \epsilon_N }\prod \limits_{k=1}^{N-1} \frac{f_{k-1}}{\mu_k + f_k + \gamma_k - \sigma_k - \epsilon_k } }, \nonumber \\
y_i = \frac{\prod \limits_{k=1}^i \frac{f_{k-1}}{\mu_k + f_k + \gamma_k - \sigma_k - \epsilon_k } }{1 + \sum \limits_{l=1}^{N-1}\prod \limits_{k=1}^l \frac{f_{k-1}}{\mu_k + f_k + \gamma_k - \sigma_k - \epsilon_k } + 
	\frac{f_{N-1}}{\mu_N + \gamma_N - \sigma_N - \epsilon_N }\prod \limits_{k=1}^{N-1} \frac{f_{k-1}}{\mu_k + f_k + \gamma_k - \sigma_k - \epsilon_k }}, \nonumber \\
i=1,\dots, N-1, \nonumber \\
y_N = \frac{\frac{f_{N-1}}{\mu_N + \gamma_N - \sigma_N - \epsilon_N }\prod \limits_{k=1}^{N-1} \frac{f_{k-1}}{\mu_k + f_k + \gamma_k - \sigma_k - \epsilon_k }}{1 + \sum \limits_{l=1}^{N-1}\prod \limits_{k=1}^l \frac{f_{k-1}}{\mu_k + f_k + \gamma_k - \sigma_k - \epsilon_k } + 
	\frac{f_{N-1}}{\mu_N + \gamma_N - \sigma_N - \epsilon_N }\prod \limits_{k=1}^{N-1} \frac{f_{k-1}}{\mu_k + f_k + \gamma_k - \sigma_k - \epsilon_k } }. \nonumber \\
\end{eqnarray}
To the best of our knowledge the class of distributions (\ref{d2}) was not discussed by other authors. 
The corresponding class of distributions for the case $N=\infty$ is discussed in Appendix A.
Several shapes of the distributions from the class (\ref{d2}) can be seen in Fig. 3 (note that the classes of distributions (\ref{d2})
and (\ref{gen_distr}) are closely connected). 
\par
We note that the system of equations (\ref{eu1}) - (\ref{eu3}) is connected to the 
system of equations
\begin{equation}\label{gen1}
x_i = F_i x_{i-1}, i=1,\dots,N
\end{equation}
where $F_i$ is a function of $i$ and eventually also a function of other variables and parameters. 
This connection can be easily proved. Let $f_0 = \mu_0 + \gamma_0 -\sigma_0 - \epsilon_0$ and we choose $f_i =\frac{f_{i-1}}{F_i} -  \mu_i  - \gamma_i + \sigma_i + \epsilon_i $; $i=1,\dots,N-1$; and $\gamma_N = \frac{f_{N-1}}{F_N} - \mu_n + \sigma_N + \epsilon_n$. Then Eq.(\ref{eu1}) is satisfied and
Eqs.(\ref{eu2}) and (\ref{eu3}) are transformed to Eq.(\ref{gen1}).

\begin{figure}[!htb]
\centering
\includegraphics[scale=.7]{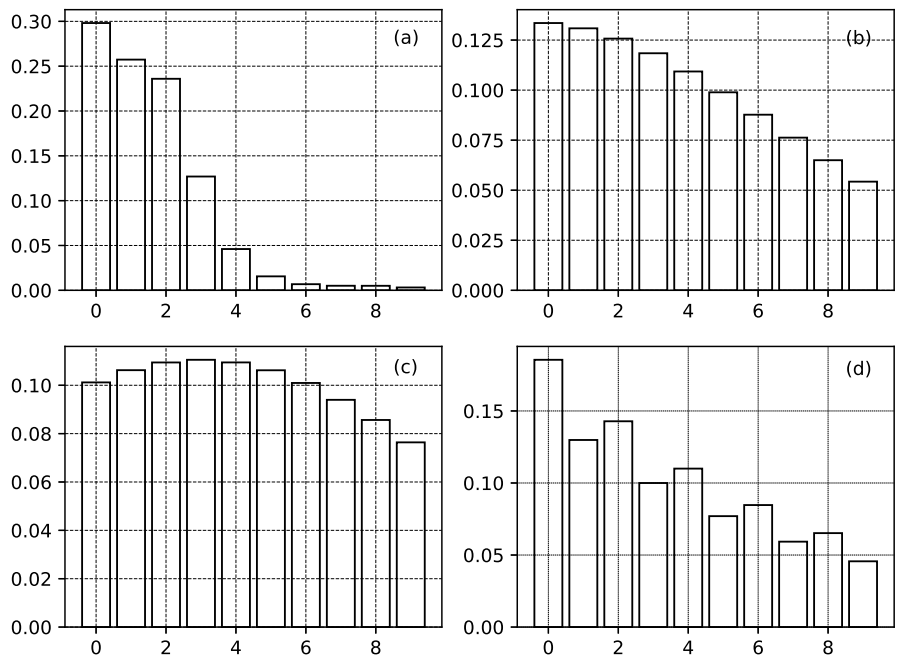}
\caption{Members of the family of truncated distributions (\ref{gen_distr}) for the case of chain
consisting of 10 urns. 
Horizontal axis - $k$. Vertical axis: probability $p(k)$. Figure (a): $F_i = 2 - \sin(i)$. 
Figure (b): $F_i = \exp(-0.02i)$ Figure (c): $F_i = 1.07 - \tanh(0.02i)$. Figure (d): 
$F_i = 0.9+0.2(-1)^i$.}
\end{figure}
\par
Eq.(\ref{gen1}) leads to a class of statistical distributions
as follows. From Eq.(\ref{gen1}) we obtain $x_i = x_0 \prod \limits_{k=1}^i F_k$ and then the 
amount of substance  in the chain of urns will be $x = x_0 \left[ 1+\sum \limits_{i=1}^N \prod \limits_{k=1}^i F_k\right]$. We can consider the statistical distribution $y_i=x_i/x$ of the 
amount of substance along the urns of the chain of urns. $y_i$ can be considered as 
probability values of distribution of a discrete random variable $\zeta$: $y_i = p(\zeta =i)$, $i=0,\dots, N$. For this distribution we obtain
\begin{equation}\label{gen_distr}
y_0 = \frac{1}{1+ \sum \limits_{l=1}^N \prod \limits_{k=1}^l F_k}; \ \ 
y_i = \frac{\prod \limits_{k=1}^i F_k}{1+\sum \limits_{l=1}^N \prod \limits_{k=1}^l F_k}, \
i=1,\dots,N
\end{equation}
\par
The distribution (\ref{gen_distr}) is closely connected to the distribution (\ref{d2}) as
$$
F_i = \frac{f_{i-1}}{\mu_i + f_i + \gamma_i - \sigma_i - \epsilon_i}; \ \ 
F_N = \frac{f_{N-1}}{\mu_N + \gamma_N - \sigma_N - \epsilon_N}.
$$
Because of the presence of the functions $F_k$ the shapes of the distributions from the class 
(\ref{gen_distr}) (and the shapes of distributions from class (\ref{d2}) respectively) can be quite different one from another. 
Several shapes are shown in Fig.2. Appendix A contains the theory for the case $N=\infty$. The  class of distributions (\ref{gen_distr}) has many interesting particular cases. In
Appendix A we show that for the case of infinite distributions the class of distributions
(generated by the equations corresponding to Eq.(\ref{d2}) or Eq.(\ref{gen_distr}) from the finite case discussed in the main text) 
contains as particular cases  the Katz family  of distributions, extended Katz family of distributions,Sundt and Jewell family of distributions, Ord family of distributions, 
and Kemp family of distributions. 
\begin{figure}[!htb]
\centering
\includegraphics[scale=.7]{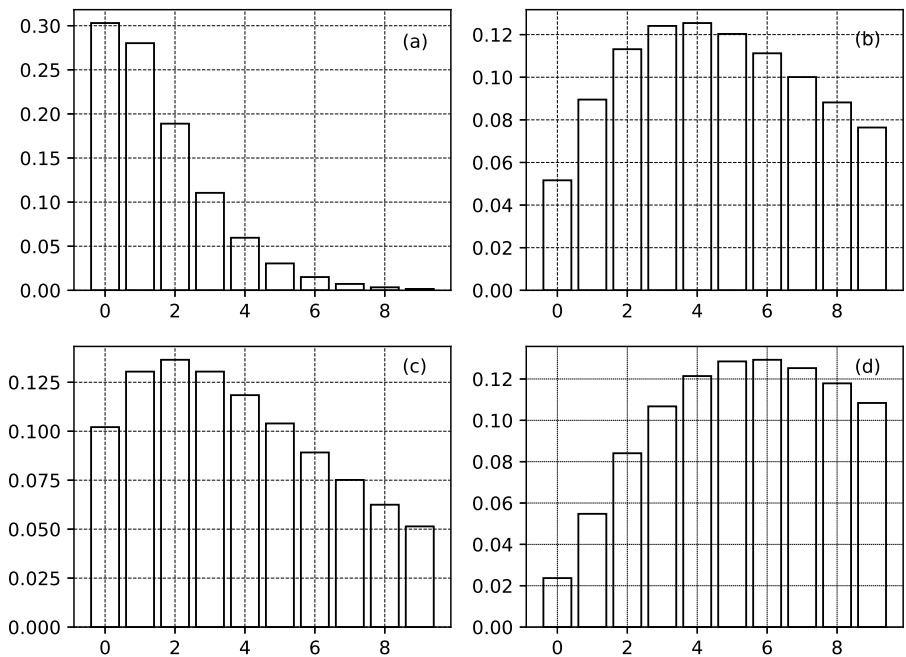}
\caption{Members of Kemp family of truncated distributions for a chain of 10 urns. Case $p=3$, $q=2$ 
in Eq.(\ref{kemp}). Horizontal axis - $k$. Vertical axis: 
probability $p(k)$. Figure (a): $\lambda=0.4$, $a_1=0.8$, $a_2 = 0.7$, $a_3=1.2$, $b_1=0.6$, $b_2=0.3$, $b_2=0.4$. Figure (b): The values of the parameters are the same as in Figure (a) except for the value 
of the parameter $\lambda$ that is $\lambda=0.75$. The effect of increasing value of $\lambda$ is
increasing the probability at the tail of  the distribution. Figure (c): The values of the parameters are the same as in Figure (b) except for the value of the parameter $b_3$ which is $b_3=0.9$. The effect of 
the increasing value of $b_1$ is opposite to the  effect of increasing value of $\lambda$ from Figure (b). Figure (d): The values of the parameters are the same as in Figure (b) except for the value of the 
parameter $a_1$ which is $a_1=1.4$.}
\end{figure}
These families contain numerous named distributions and
some of them are listed in Appendix A. Here in the main text we consider the case of finite $N$ and because of this
we obtain  families of truncated distributions. For the case of relationship corresponding to the Katz family of distributions \footnote{The original formula for the Katz family of distributions is $y_{i+1} = \frac{\alpha + \beta i}{1+i} y_i$,  $i=0, \dots$
 To follow the notation in this paper we have set $i=k-1$ and obtained (\ref{ktz1t})}
\cite{katz1},\cite{johns} the \emph{Katz family of truncated distributions}  is obtained

\begin{equation}\label{ktz1t}
y_k = \frac{\alpha+ \beta(k-1)}{k} y_{k-1}, k=1,\dots,N; \ \ 
y_0 = \frac{1}{1+ \sum \limits_{j=1}^N \prod \limits_{l=1}^j \frac{\alpha + \beta(l-1)}{l}}
\end{equation}
Above $F_k = \frac{\alpha + \beta(k-1)}{k}$, and $\alpha$ and $\beta$ are parameters.
The restrictions on the parameters are $\alpha > 0$, $\beta < 1$ (as $\beta \ge 1$ does not yield a
valid distribution). If $\alpha + \beta n < 0$, then $p_{n+j}$ is understood to be equal to zero
for all $j > 0$. 
\par 
Let $F_k = \frac{\alpha + \beta (k-1)}{\gamma + k - 1}$, $k=1,\dots,N$. Then we obtain the 
\emph{extended Katz family of truncated distributions} \cite{gurland} given by
\begin{equation}\label{ktz2t}
y_k = \frac{\alpha+ \beta(k-1)}{\gamma + k -1} y_{k-1}, k=1,\dots,N; \ \
y_0 = \frac{1}{1+ \sum \limits_{j=1}^N \prod \limits_{l=1}^j \frac{\alpha + \beta(l-1)}{\gamma+ l -1}}
\end{equation}
In Eq.(\ref{ktz2t}) $\alpha>0$, $\beta <1$ and $\gamma>0$. The resulting distribution will be named 
\emph{truncated hyper-negative binomial distribution} as the corresponding distribution for $N =\infty$
has the name \emph{hyper-negative binomial distribution} \cite{yousry}. When $\beta \to 0$ in Eq.(\ref{ktz2t}) we obtain a distribution we shall call \emph{truncated hyper-Poisson distribution}
because the corresponding distribution for the case $N = \infty$ was called
\emph{hyper-Poisson distribution} \cite{bardwell}. 
\par 
Another particular case of the distribution (\ref{gen_distr}) is the \emph{Sundt and Jewell family of  truncated 
distributions} named after the corresponding family of distributions for the case $N=\infty$ \cite{sj}, 
\cite{wilm88}. This family of distributions is obtained when
\begin{equation}\label{sundtt}
F_k = a+ \frac{b}{k}, k=1,\dots, N; \ \
y_0 = \frac{1}{1+ \sum \limits_{j=1}^N \prod \limits_{l=1}^j \left(a + \frac{b}{l} \right)}
\end{equation}
Above $a$ and $b$ are parameters.
\par
The \emph{Ord's family 
of truncated distributions} named after the Ord's family of distributions \cite{ord1} - \cite{ord3} 
is obtained when 
\begin{eqnarray}\label{ordt}
F_k &=& \frac{[(a+b_0) + (b_1-1)k + b_2k(k-1)]-(k-a)}{(a+b_0) + (b_1-1)k + b_2k(k-1)}, k =1,\dots,N \nonumber \\
y_0 &=& \frac{1}{1+ \sum \limits_{j=1}^N \prod \limits_{l=1}^j \frac{[(a+b_0) + (b_1-1)l + b_2 l (l-1)]+(l-a)}{(a+b_0) + (b_1-1)l + b_2 l (l-1)}}
\end{eqnarray}
In (\ref{ordt}) $a$ and $b_k$ are parameters. 
\par 
Finally we note that the \emph{Kemp's family of truncated distributions} named after  \emph{Kemp's family of distributions} for the case $N=\infty$ \cite{johns}, \cite{kemp1} is also a particular 
case of the the family of distributions (\ref{gen_distr}) when
\begin{eqnarray}\label{kempt}
F_k &=& \frac{\lambda (a_1+k) \dots (a_p + k)}{(b_1 + k) \dots (b_q+k)(b_{q+1}+k)}, k=1,\dots,N 
\nonumber \\
y_0 &=& \frac{1}{1+ \sum \limits_{j=1}^N \prod \limits_{l=1}^j \frac{\lambda (a_1+l) \dots (a_p + l)}{(b_1 + l) \dots (b_q+l)(b_{q+1}+l)}}
\end{eqnarray}
In (\ref{kempt}) $\lambda$, $a_i$ and $b_i$ are parameters. Several members of the Kemp family of
truncated distributions are shown in Fig. 3.
\section{Concluding remarks}
We discuss in this article a model of motion of substance in a chain of urns
with possibility for exchange of substance among the chain of urns, the surrounding network and the
environment of this network. Our attention was focused on the analytical results connected to this urn model.
Such analytical results can be obtained, e.g., for the case of stationary motion of
the substance  through the chain of urns. For the case of infinite length of the 
chain of urns  we obtain a family of discrete distributions that contains as particular cases
many other families of discrete distributions, e.g., the families of distributions of Katz, 
Ord, and Kemp (see the Appendix). We write several characteristic quantities of the obtained family of distributions,
e.g., mean and standard deviation. For the case of finite length of the chain of urns 
we obtain a family of distributions that are the truncated version of the discrete
distributions of the family of  distributions obtained for the case of chain of urns of
infinite length. In analogy to the case of infinite chain of urns we name some subfamilies
of distributions for the case of finite chain of urns as Katz, Ord, Kemp, etc. families
of truncated distributions.
\par 
The discussed urn model has different practical applications. Let us discuss briefly just one
of these applications: motion of migrants through a migration channel. In this case the chain 
of urns corresponds to the chain of countries that form the migration channel. The substance
 that is presented in the urns and moves between them corresponds to the migrants that
move between the countries of the migration channel. The pumping process corresponds to
inflow of migrants in the countries of the channel. This inflow can come from the network (countries
from the same continent where the migration channel is positioned) or from the environment of
the network (corresponding to the countries from the other continents). The process of leakage corresponds to outflow of migrants from the channel. This outflow can be to the countries from the 
continent where the channel is positioned (leakage to the network). The outflow can be also 
to the countries from other continents (outflow to the environment of the network). The obtained family of distributions 
corresponds to the distribution of migrants along the countries of the migration channel for the case of  
stationary regime of motion of migrants in the migration channel.
\begin{appendix}
\section{Statistical distributions of substance for the case of chain of urns
containing infinite number of urns}
\par 
The model of infinite chain of urns corresponding to the model of finite chain of urns described by by Eqs.(\ref{ez1}) - (\ref{ez3}) is 
\begin{eqnarray}\label{ez1x}
\bigg[ \sigma_0(t_k) - \mu_0(t_k) - f_0(t_k)  -  \gamma_0(t_k) + \epsilon_0(t_k) \bigg] x_0(t_k) = 0,
\end{eqnarray}
\begin{eqnarray}\label{ez2x}
\bigg[ \mu_i(t_k) + f_i(t_k) + \gamma_i(t_k) - \sigma_i(t_k) - \epsilon_i(t_k) \bigg] x_i(t_k) = f_{i-1}(t_k) x_{i-1}(t_k),
\nonumber \\
i=1,\dots, \infty.
\end{eqnarray}
Let us consider  the case when the parameters of the model are time independent
(i.e., when $\sigma_i(t_k) = \sigma$; $\mu_i(t_k) = \alpha_i$, $i=0,\dots,N$;
$\gamma_i(t_k) = \gamma_i$; $\epsilon_i(t_k) = \epsilon_i$; $f_0(t_k)=f_0$;
$i=0,\dots,N$). In this case the system of model equations becomes
\begin{eqnarray}\label{eu1x} 
\bigg[ \sigma_0 - \mu_0 - f_0  -  \gamma_0 + \epsilon_0 \bigg] x_0 = 0
\end{eqnarray}
\begin{eqnarray}\label{eu2x}
\bigg[ \mu_i + f_i + \gamma_i - \sigma_i - \epsilon_i \bigg] x_i = f_{i-1} x_{i-1}, \ \
i=1,\dots,\infty
\end{eqnarray}
From Eqs. (\ref{eu1x}) and (\ref{eu2x}) we obtain the relationships
\begin{eqnarray}\label{d1y}
	f_0 &=& \sigma_0 - \mu_0  -  \gamma_0 + \epsilon_0; \nonumber \\
	x_i &=& x_0 \prod \limits_{k=1}^i \frac{f_{k-1}}{\mu_k + f_k + \gamma_k - \sigma_k - \epsilon_k },
	\nonumber \\
\end{eqnarray}
From Eqs. (\ref{d1y})  we obtain 
\begin{eqnarray}\label{d1x}
x &=& \sum \limits_{l=0}^\infty x_l = x_0 \Bigg[ 1 + \sum \limits_{l=1}^{\infty}\prod \limits_{k=1}^l \frac{f_{k-1}}{\mu_k + f_k + \gamma_k - \sigma_k - \epsilon_k }  \Bigg]
\end{eqnarray}
We can consider the statistical distribution $y_i=x_i/x$ of the 
amount of substance along the urns of the sequence of urns. $y_i$ can be considered as 
probability values of distribution of a discrete random variable $\zeta$: $y_i = p(\zeta =i)$, $i=1,\dots, \infty$. 
For this distribution we obtain
\begin{eqnarray}\label{d2x}
y_0 &=&  \frac{1}{1 + \sum \limits_{l=1}^{\infty}\prod \limits_{k=1}^l \frac{f_{k-1}}{\mu_k + f_k + \gamma_k - \sigma_k - \epsilon_k }} \nonumber \\
y_i &=& \frac{\prod \limits_{k=1}^i \frac{f_{k-1}}{\mu_k + f_k + \gamma_k - \sigma_k - \epsilon_k } }{1 + \sum \limits_{l=1}^{\infty}\prod \limits_{k=1}^l \frac{f_{k-1}}{\mu_k + f_k + \gamma_k - \sigma_k - \epsilon_k }}, \ \ i=1,\dots, \infty 
\end{eqnarray}
To the best of our knowledge the class of distributions (\ref{d2x}) was not discussed by other authors. 
\par
Note that the above parameters don't depend on the time but they can depend  on
$i$ and on other parameters. Eqs. (\ref{eu1x}) and (\ref{eu2x}) are connected to the equation   
\begin{equation}\label{genx1}
x_i = F_i x_{i-1}, i=1,\dots, \infty
\end{equation}
where $F_i$ is a function of $i$ and eventually also a function of other variables and parameters. 
This connection can be easily shown. Let $f_0 = \mu_0 + \gamma_0 -\sigma_0 - \epsilon_0$; $f_i =\frac{f_{i-1}}{F_i} - \mu_i  - \gamma_i + \sigma_i + \epsilon_i $; \ $i=1,\dots,$. Then Eq.(\ref{eu1x}) is satisfied and Eq.(\ref{eu2x}) is transformed to Eq.(\ref{genx1}).
Eq.(\ref{genx1}) leads to a class of statistical distributions
as follows. From Eq.(\ref{genx1}) we obtain $x_i = x_0 \prod \limits_{k=1}^i F_k$ and then the 
amount of substance  in the chain of urns will be $x = x_0 \left[ 1+\sum \limits_{l=1}^\infty \prod \limits_{k=1}^l F_k\right]$. 

We can consider the statistical distribution $y_i=x_i/x$ of the 
amount of substance along the urns of the sequence of urns. $y_i$ can be considered as 
probability values of distribution of a discrete random variable $\zeta$: $y_i = p(\zeta =i)$, $i=1,\dots, \infty$. 
For this distribution we obtain
\begin{equation}\label{gen_distrx}
y_0 = \frac{1}{1+ \sum \limits_{l=1}^\infty \prod \limits_{k=1}^l F_k}; \ \ 
y_i = \frac{\prod \limits_{k=1}^i F_k}{1+\sum \limits_{l=1}^\infty \prod \limits_{k=1}^l F_k}, \
i=1,\dots, \infty
\end{equation}
\par
We can write Eq.(\ref{genx1}) as follows
\begin{equation}\label{gd1x}
y_i = F_i y_{i-1}
\end{equation}
(Note that $F_i = \frac{f_{i-1}}{\mu_i+f_i+\gamma_i - \sigma_i - \epsilon_i}$ for the case of distribution (\ref{d2x})).
Several members of the family of distributions (\ref{gen_distrx}) are shown in Fig. 10.
\begin{figure}[!htb]
\centering
\includegraphics[scale=.7]{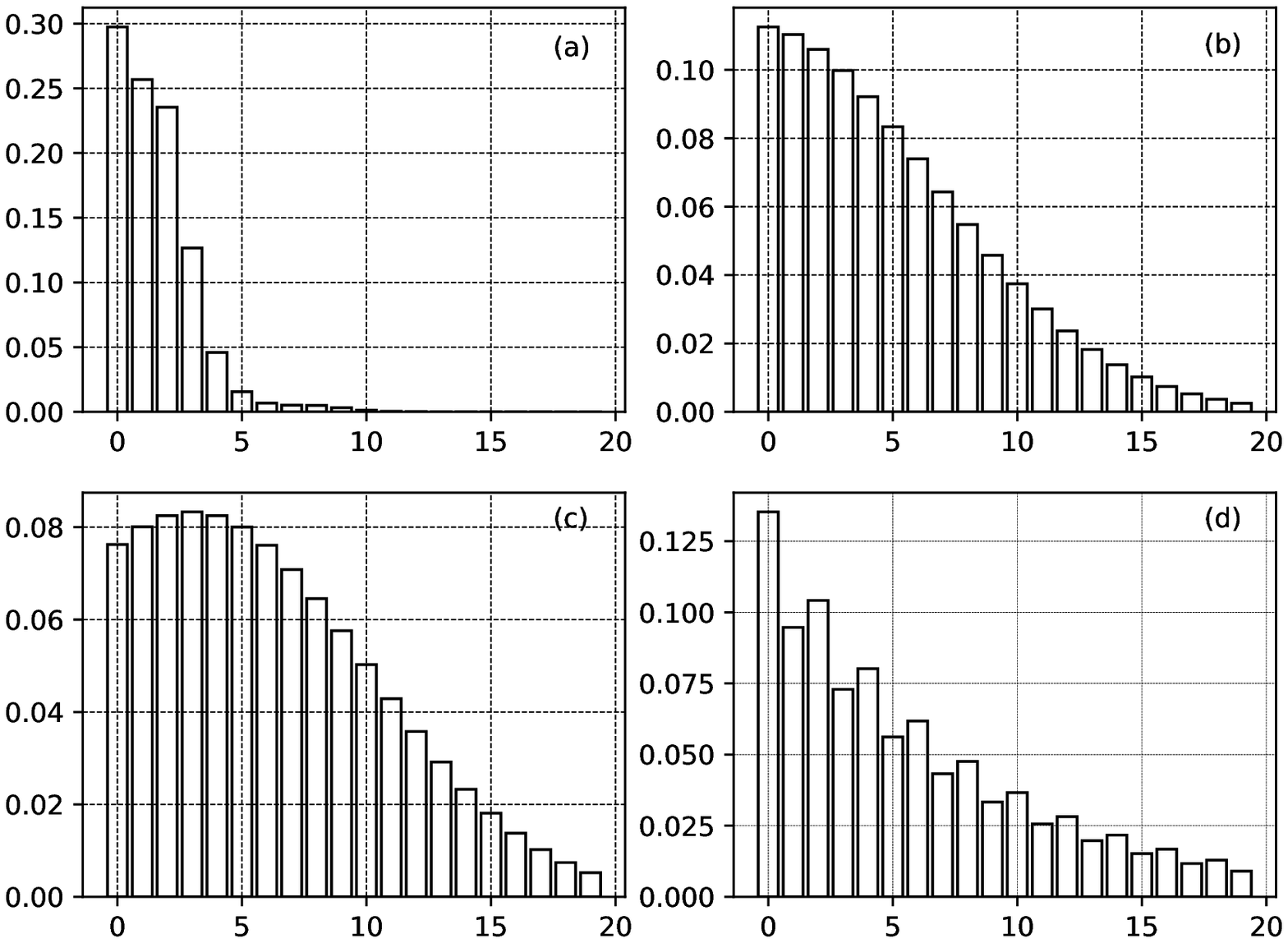}
\caption{Members of the family of distributions (\ref{gen_distrx}). 
Horizontal axis - $k$. Vertical axis: probability $p(k)$. Figure (a): $F_i = 2 - \sin(i)$. 
Figure (b): $F_i = \exp(-0.02i)$ Figure (c): $F_i = 1.07 - \tanh(0.02i)$. Figure (d): 
$F_i = 0.9+0.2(-1)^i$.}
\end{figure}
Particular case of the distributions defined by Eq.(\ref{gd1x}) is the Katz family of distributions
\footnote{The original formula for the Katz family of distributions is $y_{i+1} = \frac{\alpha + \beta i}{1+i} y_i$, where $i=0, \dots$. In order to follow the notation in this paper we have set $i=k-1$ and thus (\ref{ktz1}) is obtained}
 \cite{katz1}
\begin{equation}\label{ktz1}
y_k = \frac{\alpha+ \beta(k-1)}{k} y_{k-1}, \ \  k=1,\dots
\end{equation}
Above $F_k = \frac{\alpha + \beta(k-1)}{k}$, and $\alpha$ and $\beta$ are parameters.
The restrictions on the parameters are $\alpha > 0$, $\beta < 1$ (as $\beta \ge 1$ does not yield a
valid distribution). If $\alpha + \beta n < 0$, then $p_{n+j}$ is understood to be equal to zero
for all $j > 0$. The Katz family of distributions contains the following important distributions
\cite{katz1}, \cite{johns}:
\begin{itemize}
\item \emph{Binomial distribution :} For $\beta < 0$ one obtains  the binomial distribution with 
$\alpha = \frac{nq}{p}$, $\beta = - \frac{q}{p}$,
\item \emph{Poisson distribution:} It is obtained when $\beta = 0$ and
$\alpha = \theta$,
\item \emph{Negative binomial distribution:} It is obtained when $ 0 < \beta <1$
and $\alpha = \frac{jP}{P+1}$; $\beta = \frac{P}{P+1}$.
\end{itemize}
The three distributions above are particular cases of the class of distributions defined
by Eq.(\ref{gd1x}).
\par 
Let $F_k = \frac{\alpha + \beta (k-1)}{\gamma + k - 1}$, $k=1,\dots$. Then we obtain the 
\emph{extended Katz family of distributions} \cite{gurland} given by
\begin{equation}\label{ktz2}
y_k = \frac{\alpha+ \beta(k-1)}{\gamma + k -1} y_{k-1},\ \  k=1,\dots
\end{equation}
In Eq.(\ref{ktz2}) $\alpha>0$, $\beta <1$ and $\gamma>0$. The resulting distribution was called
\emph{hyper-negative binomial distribution} \cite{yousry}. When $\beta \to 0$ in Eq.(\ref{ktz2})
one obtains the \emph{hyper-Poisson distribution} \cite{bardwell}. Several members of the Katz family of distributions and extended Katz family of distributions are shown in Fig.11.
\begin{figure}[!htb]
\centering
\includegraphics[scale=.7]{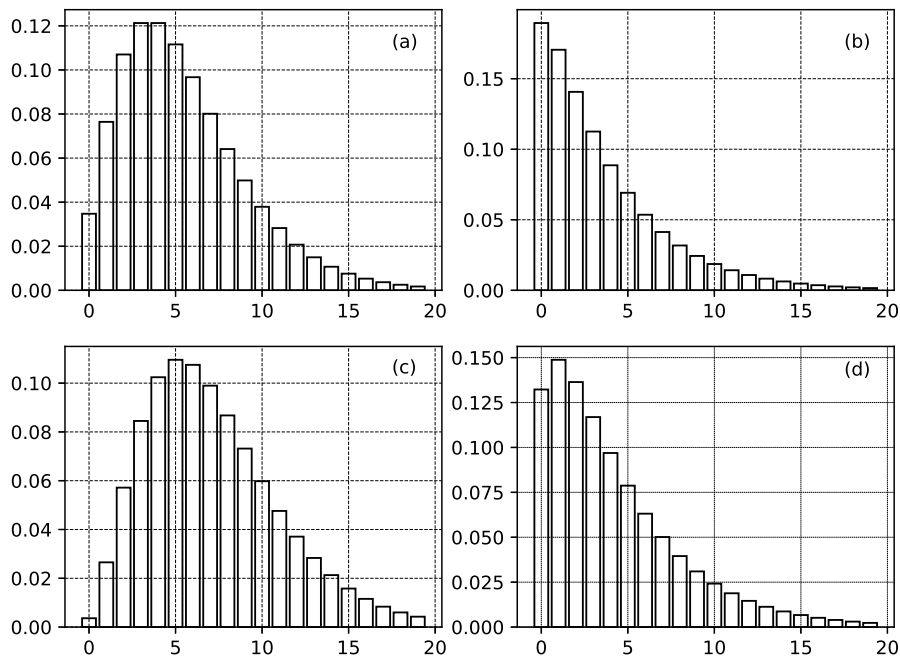}
\caption{Members of Katz family of distributions (Figures (a) and (b)) and of extended Katz
families of distributions (Figures (c) and (d)). Horizontal axis - $k$. Vertical axis: 
probability $p(k)$. Figure (a): $\alpha = 2.2$, $\beta=0.6$. Figure (b):
$\alpha=0.9$, $\beta=0.75$. Figure (c): a member of the extended Katz family of distributions. The values 
of $\alpha$ and $\beta$ are the same as in Figure (a). The additional parameter $\gamma$ of the extended 
Katz family of distributions has value $\gamma=0.3$. The presence of the 
parameter $\gamma$ leads to increasing probability in the tail of the distribution and to change of the 
shape of distribution. Figure (d): another member of the extended Katz family of distributions. Values of 
parameters $\alpha$ and $\beta$ are the same as in Figure (b). The value of parameter $\gamma$ is
$\gamma=0.8$.}
\end{figure}
\par 
Another particular case of (\ref{gd1x}) is the \emph{Sundt and Jewell family of distributions}. It is obtained when \cite{sj}, \cite{wilm88}
\begin{equation}\label{sundt}
F_k = a+ \frac{b}{k}, k=1,\dots.
\end{equation}
The \emph{Ord's family of distributions} \cite{ord1} - \cite{ord3} is obtained when 
\begin{equation}\label{ord}
F_k = \frac{[(a+b_0) + (b_1-1)k + b_2k(k-1)]- (k-a)}{(a+b_0) + (b_1-1)k + b_2k(k-1)}.
\end{equation}
In (\ref{ord}) $a$ and $b_k$ are parameters.
Ord obtained the following kinds of distributions:
\emph{hypergeometric}, \emph{negative hyper-geometric (beta-binomial)}, \emph{beta-Pascal}, \emph{
binomial}, \emph{Poisson}, \emph{negative binomial}, \emph{discrete Student's t}. All these case of distributions from the Ord's family of distributions are also particular cases of the lass of distributions (\ref{gen_distrx}). 
\par 
Finally we note that the Kemp's family of distributions (the generalized hypergeometric
probability distributions) \cite{johns}, \cite{kemp1} is also a particular 
case of the the family of distributions (\ref{gen_distrx}) when
\begin{equation}\label{kemp}
F_k = \frac{\lambda (a_1+k) \dots (a_p + k)}{(b_1 + k) \dots (b_q+k)(b_{q+1}+k)}.
\end{equation}
In (\ref{kemp}) $\lambda$, $a_i$ and $b_i$ are parameters.
Dacey \cite{dacey} (see Table 2 there) lists 41 named distributions that belong to the Kemp's family 
of distributions. Let us mention some of them as they are also particular cases of the class of
distributions (\ref{gen_distrx}) (or (\ref{d2x})). The distributions we mention here are: \emph{Binomial distribution}, \emph{Generalized Poisson Binomial distribution}, \emph{Poisson distribution}, \emph{Hyper-Poisson distribution}, 
\emph{Geometric distribution}, \emph{Pascal distribution}, \emph{Negative Binomial distribution},\
\emph{Polya-Eggenberger distribution}, \emph{Stirling distributions of first and second kind},
\emph{Hypergeometric, Inverse Hypergeometric and Negative Hypergeometric distributions}, \emph{Polya and Inverse Polya distributions}, \emph{Hermite distribution}, \emph{Neuman distribution of type A, B, 
and C}, \emph{Waring distribution}, \emph{Yule distribution}.

\end{appendix}

\end{document}